\documentclass[twocolumn,prl,a4paper,floatfix,preprintnumbers,amsmath,amssymb]{revtex4}
\usepackage{graphicx}
\begin{document}

\title{Phase Estimation With Interfering Bose-Condensed Atomic Clouds}

\author{Jan Chwede\'nczuk$^{1,2}$, Francesco Piazza$^{1,2}$ and Augusto Smerzi$^{1,2}$}

\affiliation
    {
      $^1$INO-CNR, BEC Center, Via Sommarive 14, 38123 Povo, Trento, Italy\\
      $^2$Dipartimento di Fisica, Universita' di Trento, 38123 Povo, Italy
    }

\begin{abstract} 
  We investigate how to estimate from atom-position measurements the relative phase of two Bose-Einstein condensates
  released from a double-well potential. We demonstrate that the phase estimation sensitivity via the fit of the average
  density to the interference pattern is fundamentally bounded by shot noise. This bound can be overcome
  by estimating the phase from the measurement of $\sqrt N$ (or higher) correlation function. 
  The optimal estimation strategy requires the measurement of the $N$-th order correlation function. 
  We also demonstrate that a second estimation method -- based on the detection of the center of mass 
  of the interference pattern -- provides sub shot-noise sensitivity. 
  Yet, the implementation of both protocols might be experimentally challenging.
\end{abstract}

\maketitle

{\it Introduction}. 
The experimental realization of the Bose-Einstein condensation (BEC) opened a new chapter in the field of 
atom interferometry \cite{cronin}. The BEC constitutes a bright
and well-controllable source of particles, in analogy to the laser light commonly
used for interferometric purposes. Atoms are very good candidates
for precise measurement of electromagnetic \cite{battesi, cadoret, harber} or
gravitational \cite{peters, lamporesi} forces. Moreover, non-classical states have been created via Quantum Nondemolition measurements \cite{appel, smith}, or exploiting the naturally present two-body interactions \cite{esteve,riedel, maussang}. These states allow to overcome the limit imposed by the classical physics on measurement precision (shot-noise limit) \cite{giovanetti, pezze}, as recently demonstrated experimentally \cite{gross}. 
An atom interferometer can be implemented using a gas trapped in a double-well potential \cite{shin, schumm, wang
}, where the two 
spatial modes localized in its minima play the role of the interferometer's arms. 
In the simplest interferometric sequence, first
a relative phase $\theta$ builds
up between the wells and then the BECs are released from the trap and form an interference pattern. 
In this manuscript we analyze different phase estimation strategies relying on position measurement of the atoms
forming this pattern. We show that fitting the average density to the interference pattern \cite{shin}, 
gives a sensitivity $\Delta\theta$ bounded by the shot-noise.
This limit can be overcome by measuring correlations between the atoms of order not smaller than $\sqrt N$, where
$N$ is the total number of atoms. We also
demonstrate that the
estimation by the $N$-th order correlation function is {\it optimal} -- 
the sensitivity saturates the bound set by the Quantum Fisher Information (QFI) \cite{braun}. Finally,
we present another estimation scheme which can give sub shot-noise sensitivity, based on the measurement of
the position of the center of mass of the cloud. Although it involves a probability which is simpler
to construct with respect to high-order correlation functions, it requires detection of all $N$ atoms. These results 
indicate that the achievement of sub shot-noise sensitivity using the 
interference pattern might prove challenging.

{\it The model}. 
We introduce the two-mode field operator of a bosonic gas in a
double-well potential, $\hat\Psi(x,t)=\psi_a(x,t)\hat a+\psi_b(x,t)\hat b$, 
where $\hat a^\dagger/\hat b^\dagger$ creates
an atom in the left/right well. While the atoms remain trapped, an unknown relative phase $\theta$ 
is imprinted between the wells.
This stage is described by a unitary evolution $\hat U(\theta)\!\!=\!\!e^{-i\theta\hat J_z}$ 
of the initial state $|\psi_{in}\rangle$ of the double-well system \cite{nota_J}.
After the phase acquisition, the trap is switched off and the initially localized mode functions $\psi_{a/b}(x,t)$ 
freely expand.

All information which can be extracted from a position measurement is contained in $p_N(\vec x_N|\theta)$ \cite{javanainen, castin, laloe_old}--  
the conditional probability of finding $N$ particles at positions $\vec x_N=(x_1\ldots x_N)$. It can be 
expressed in terms of the $N$-th order correlation 
function \cite{nota_corr} as follows: $p_N(\vec x_N|\theta)=\frac1{N!}G_N(\vec x_N,\theta)$.
To evaluate this probability, we write the initial state in the well-population basis, 
$|\psi_{in}\rangle=\sum_{n=0}^NC_n|n,N\!\!-\!n\rangle$ and suppose
that the expansion coefficients are real and posses the symmetry $C_n=C_{N-n}$ \cite{nota_cn}. It is convenient to use
the Heisenberg representation and evolve the field operator, 
$\hat U^\dagger(\theta)\hat\Psi(x,t)\hat U(\theta)=\psi_a(x,t)e^{i\frac\theta2}\hat a+\psi_b(x,t)e^{-i\frac\theta2}\hat b$.
Then, we expand the Fock states $|n,N\!-\!n\rangle$ in the
basis of the spin-coherent states \cite{laloe} and obtain
\begin{eqnarray}\label{prob_ph}
  &&p_N(\vec x_N|\theta)=
  \int\limits_0^{2\pi}\int\limits_0^{2\pi}\frac{d\varphi}{2\pi}\frac{d\varphi'}{2\pi}\prod_{i=1}^Nu^*_\theta(x_i,\varphi;t)
  u_\theta(x_i,\varphi';t)\nonumber\\
  &&\sum_{n,m=0}^N\frac{C_nC_m\cos\left[\varphi\left(\frac N2-n\right)\right]
    \cos\left[\varphi'\left(\frac N2-m\right)\right]}{\sqrt{\binom{N}{n}\binom{N}{m}}},
\end{eqnarray}
where $u_\theta(x,\varphi;t)=\psi_a(x,t)e^{\frac i2(\varphi+\theta)}+\psi_b(x,t)e^{-\frac i2(\varphi+\theta)}$.
In the remaining part of the manuscript, the time $t$ is chosen such that the interference pattern is already formed. 
The probability (\ref{prob_ph}) is the starting point for the following discussion of various phase estimation strategies.

{\it Estimation via the fit to the density}.
One of the most natural ways of estimating the value of $\theta$ is through the fit of 
the average density to the interference pattern. The experiment consists in dividing
the interference pattern into $M$ bins located at positions $x_i$ 
and measuring the number of particles $n_i$ in each of them. This procedure
is repeated $m$ times: the set $n_i^{(k)}\!\!,$ $k=1,...,m$ is recorded 
and the average $\bar{n}_i=\sum_{k=1}^mn_i^{(k)}/m$ in $i$-th bin is evaluated. 
The average occupation $\langle n_i\rangle=\lim_{m\to\infty}\sum_{k=1}^mn_i^{(k)}/m$ 
(we assume to be known a priori in the experiment) with free parameter $\theta$ is then fitted to the histogram
of the measured density $\{x_i, \bar{n}_i\}, i=1,...,M$. 
For small bin size $\Delta x$, $\langle n_i\rangle$
is related to the average density by 
$\langle n_i\rangle=G_1(x_i,\theta)\Delta x$.
The value of $\theta$ can be determined from the least squares formula 
$\frac{d}{d\theta}\sum_{i=1}^M\frac{(\bar{n}_i-\langle n_i\rangle)^2}{2\Delta^2 n_i/m}=0$. 
The number fluctuations $\Delta^2 n_i=\lim_{m\to\infty}\sum_{k=1}^m(n_i^{(k)}-\langle n_i\rangle)^2$ (also assumed to be known a priori in the experiment) can be calculated 
from the probability
$p(n_i|\theta)$ of detecting $n_i$ particles in the $i$-th bin,
\begin{equation*}
  p(n_i|\theta)=\binom{N}{n_i}\int\limits_{\Delta x_i}\!\!d\vec x_{n_i}\!\!\!\!\!
  \int\limits_{\mathbb{R}-\Delta x_i}\!\!\!\!d\vec x_{N-n_i}\ p_N(\vec x_N|\theta).  
\end{equation*}
Using Eq.(\ref{prob_ph}) we obtain
\begin{equation}\label{fluctuations}
  \Delta^2n_i=G_1(x_i,\theta)\Delta x+\left[G_2(x_i,x_i,\theta)-G^2_1(x_i,\theta)\right](\Delta x)^2.
\end{equation}

In order to derive
the sensitivity $\Delta^2\theta$ for this estimation method, we notice that, if the number of measurements $m$ is large, according to the central limit theorem the
probability distribution for the average $\bar{n}_i$ in the $i$-th bin tends to the Gaussian
$p(\bar{n}_i|\theta)=\frac{1}{\sqrt{2\pi}\Delta n_i/\sqrt m}e^{-\frac{(\bar{n}_i-\langle n_i\rangle)^2}{2\Delta^2 n_i/m}}$. 
Although in every shot the atom counts are correlated between the bins, averaging over large number
of experiments washes out this dependence. Therefore, the total probability of measuring
the values 
$\{\bar{n}\}=(\bar{n}_1\ldots \bar{n}_M)$ is a product 
$P(\{\bar{n}\}|\theta)=\prod_{i=1}^Mp(\bar{n}_i|\theta)$. For this Gaussian probability, 
the condition for the least square fit coincides with the condition for the maximum 
likelihood estimator (MLE), $\frac{d}{d\theta}P(\{\bar{n}\}|\theta)=0$. 
This is a crucial observation since it is well known \cite{crlb} that, for this choice of the estimator,
the sensitivity is given by the Cramer-Rao Lower Bound, $\Delta^2\theta=F^{-1}$. Here, $F$ is the Fisher information (FI),
\begin{eqnarray}\label{ml}
  &&F=\sum_{\bar n_1\ldots \bar n_M=0}^N\frac{1}{P(\{\bar{n}\}|\theta)}
  \left(\frac{\partial}{\partial\theta}P(\{\bar{n}\}|\theta)\right)^2\nonumber\\
  &&\xrightarrow{m\gg1}m\sum_{i=1}^M\frac{1}{\Delta^2n_i}\left(\frac{\partial\langle n_i\rangle}{\partial\theta}\right)^2.
\end{eqnarray}
Since conditions for the MLE and the fit coincide,
the sensitivity of the latter is given by the inverse of
(\ref{ml}) as well. 

We now demonstrate that this sensitivity is bounded by
the shot-noise. Let us assume for the moment 
that the second term in the Eq.(\ref{fluctuations}) -- which is proportional to $(\Delta x)^2$ -- can be neglected. 
If this is the case, the particle number distribution is Poissonian and the FI from (\ref{ml}) simplifies to
\begin{eqnarray}\label{fish}
  &&F=m\sum_{i=1}^M\frac{1}{G_1(x_i,\theta)}\left(\frac{\partial}
  {\partial\theta}G_1(x_i,\theta)\right)^2\Delta x\nonumber\\
  &&\simeq mN\!\!\!\int\limits_{-\infty}^\infty\! dx\frac{1}{p_1(x|\theta)}\left(\frac{\partial}
  {\partial\theta}p_1(x|\theta)\right)^2,
\end{eqnarray}
with the one-particle probability $p_1(x|\theta)=\frac{1}{2}(|\psi_a(x,t)|^2+|\psi_b(x,t)|^2)
+\frac2N\langle\hat J_x\rangle\mathrm{Re}\left[\psi_a^*(x,t)\psi_b(x,t)e^{i\theta}\right]$. It can be demonstrated
\cite{long} that Eq.(\ref{fish}) gives $F\leq mN$, 
thus $\Delta\theta\geq\Delta\theta_{SN}=\frac{1}{\sqrt{m}}\frac{1}{\sqrt{N}}$ for {\it any} input state 
($\Delta\theta_{SN}$ denotes the shot-noise sensitivity). 
Below we argue that the inclusion of the second term in the fluctuations does not improve the sensitivity. 

In Eq.(\ref{fluctuations}), the first term  $G_1(x_i,\theta)\Delta x$ scales linearly with $N$ while
the second term, as a function of $N$, is a polynomial of the order not higher than two,
$aN^2+bN+c$. Since the fluctuations $\Delta^2n_i$ must be positive, then $a\geq0$. 
Otherwise, for large $N$, no matter how small $\Delta x$, we would have $\Delta^2n_i<0$. 
If $a>0$, the positive second term enlarges the fluctuations and thus worsens the sensitivity. 
When $a=0$, the 
first and second terms in Eq.(\ref{fluctuations}) scale linearly with $N$, and for small $\Delta x$ the second term can be neglected, thus
we end up again with Eq.(\ref{fish}) (a similar argument can be made in order to conclude that increasing the bin size $\Delta x$ always worsens the sensitivity with respect to Eq.(\ref{fish})).
This is the first important result of this manuscript: $\Delta\theta_{SN}$
is a lower bound for the fit sensitivity.
This is because the maximal value of the
FI 
is expressed in terms of the single-particle probability and does not exploit the correlations 
between the atoms. Let us now demonstrate that the measurement of these correlations can indeed improve the phase sensitivity.

{\it Estimation via the correlation functions}.
The estimation method we discuss now relies upon determination of the phase $\theta$ from the
$k$-th order correlation function 
$G_k(\vec x_k|\theta)$. We can choose to deduce $\theta$ using the MLE \cite{nota_MLE}:
given a $k$-tuple $\vec \xi_k$ of detected atoms' positions, the estimated value of the phase is chosen by maximizing $G_k(\vec \xi_k|\theta)$ with respect to $\theta$.
The variance $\Delta^2\theta$ after $m\gg1$ experiments is given by $\Delta^2\theta=F_{(k)}^{-1}$, where 
\begin{equation}\label{fish_m}
  F_{(k)}=m\frac Nk\int\limits_{-\infty}^\infty\! d\vec x_k\frac{1}
  {p_k(\vec x_k|\theta)}\left(\frac{\partial}{\partial\theta}p_k(\vec x_k|\theta)\right)^2,
\end{equation}
with $p_k(\vec x_k|\theta)=\frac{(N-k)!}{N!}G_k(\vec x_k,\theta)$. The coefficient $\frac{N}{k}$ accounts for the number
of independent drawings of $k$ particles from $N$, i.e. $\binom{N}{k}/\binom{N-1}{k-1}$. We notice 
that by setting $k=1$, i.e. the estimator is a single-particle density,
we recover the FI from Eq.(\ref{fish}). Therefore, the
measurement of positions of $N$ particles independently is, in terms of sensitivity, 
equivalent to fitting the average density, and thus is limited by the shot-noise.

\begin{figure}
  \includegraphics[scale=.36,clip]{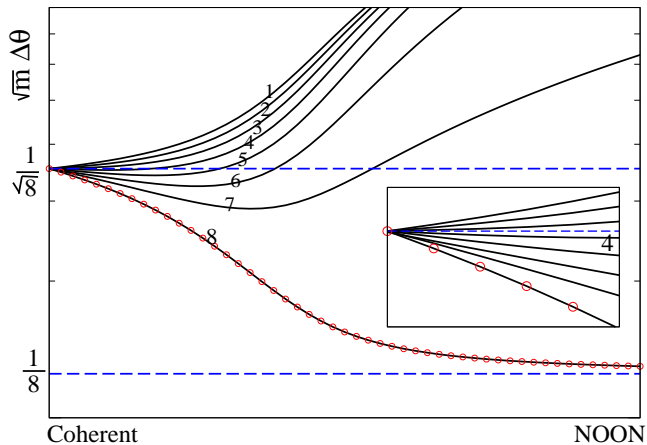}
  \caption
      {
        (color online). The sensitivity $\sqrt m\Delta\theta$ (black solid lines) for $N=8$
        calculated by numerical integration of Eq.(\ref{fish_m}) for various $k$ as a function of 
        $|\psi_{in}\rangle\in\mathcal{A}$. The two limits,
        $\sqrt m\Delta\theta_{SN}$ and $\sqrt m\Delta\theta_{HL}$ are
        denoted by the upper and lower dashed blue lines, respectively. The optimal sensitivity, given by the
        QFI, is drawn with red open circles. 
        The inset magnifies the vicinity of the spin-coherent state, showing that the sub shot-noise sensitivity is reached 
        starting from $k_{\mathrm{min}}=4$. 
      }
      \label{corr}
\end{figure}

Let us calculate the FI for the case $k=N$. 
As the interference pattern is formed after a long expansion time,
the mode functions can be written as
$\psi_{a/b}(x,t)\simeq e^{i\frac{x^2}{2\tilde\sigma^2}\mp i\frac{x\cdot x_0}{\tilde\sigma^2}}\cdot
\tilde\psi\left(\frac x{\tilde\sigma^2}\right)$,
where $\tilde\sigma=\sqrt{\frac{\hbar t}{m}}$, $\tilde\psi$ is a Fourier transform of the initial wave-packets (common
for $\psi_a$ and $\psi_b$) and the separation of the wells is $2x_0$. The above mode functions
give the probability (\ref{prob_ph}), which is inserted into 
Eq.(\ref{fish_m}). After integration over space we get
\begin{equation}\label{fish_q}
  F_{(k=N)}=m\cdot4\sum_{n=0}^NC_n^2\left(n-\frac N2\right)^2=m\cdot4\Delta^2\hat J_z.
\end{equation}
This is the second important result of this manuscript: 
the estimation of $\theta$ from the $N$-th order correlation function
is {\it optimal} -- it saturates the 
QFI, which is a result of maximization of the FI with respect to all
possible measurements \cite{braun,nota_qfi}. According to Eq.(\ref{fish_q}), all states for which
$\Delta^2\hat J_z>\frac N4$ give $\Delta\theta<\Delta\theta_{SN}$. 
Looking for sub shot-noise sensitivity, we take a sub-set of these states -- a family of
ground states of a double-well system with negative interactions, as this is a natural choice in the context of this work. 
These states range from spin-coherent to the NOON state. 
For all states in this family, which we denote by $\mathcal{A}$, the estimation by the measurement
of the $N$-th order correlation function gives  $\Delta\theta<\Delta\theta_{SN}$, and reaches
the Heisenberg limit $\Delta\theta_{HL}=\frac1{\sqrt m}\frac1N$ for the NOON state (as indicated by open circles in
Fig. \ref{corr}). 

It is important to know whether it is still possible to achieve sub shot-noise sensitivity by using lower order correlation functions $k<N$. 
As the space integrals in Eq.(\ref{fish_m}) for $k\neq N$ cannot be evaluated analytically, 
we calculate $F$ numerically taking Gaussian wave-packets, 
$\tilde\psi\left(\frac{x}{\tilde\sigma^2}\right)=
\left(\frac{2\sigma_0^2}{\pi\tilde\sigma^4}\right)
^{\frac14}e^{-\frac{x^2\sigma_0^2}{\tilde\sigma^4}}$ with the initial width of the wave-packets $\sigma_0=0.1$ and 
half of the well separation $x_0=1$. 
In Fig. \ref{corr} we plot the sensitivity using Eq.(\ref{fish_m}) for $N=8$ atoms as a function of 
$|\psi_{in}\rangle\in\mathcal{A}$.
We observe that the sensitivity improves with growing $k$ and overcomes the shot-noise limit at $k_{\mathrm{min}}=4$.

Moreover, we numerically checked that, for larger $N$, $k_{\mathrm{min}}$ tends to $\sqrt N$ (more precisely to the closest integer). 
However, measuring such a (typically) high order correlation function, during the required calibration stage ( $p_k$ must be known before performing the phase estimation protocol described above), involves a large experimental effort. 
Indeed, for every $\theta$, the $k_{\mathrm{min}}$-dimensional
function $p_{k_{\mathrm{min}}}(\vec x_{k_{\mathrm{min}}},\theta)$ must be probed. 
In the following section, we present a different detection scheme which gives sub shot-noise sensitivity and
does not require the knowledge of a multi-dimensional function.

{\it Estimation via the center of mass measurement}.
Estimation of $\theta$ from the measurement of the center of mass requires a relatively simple
calibration stage.
Positions of $N$ atoms are recorded and from this data location of the center of mass is calculated.
Such observable is described by the one-dimensional probability $p_{cm}(x|\theta)$,
related to the full $N$-body probability (\ref{prob_ph}) by
\begin{equation*}
  p_{cm}(x|\theta)=\int d\vec x_N\,\delta\left(x-\frac1N\sum_{i=1}^Nx_i\right)p_N(\vec x_N|\theta),
\end{equation*}
where ``$\delta$'' is the Dirac delta. 
Modeling the mode-functions by Gaussians, as in previous section,
and under the realistic assumption $e^{-x_0^2/\sigma_0^2}\ll1$, we obtain
\begin{equation*}
  p_{cm}(x|\theta)\propto e^{-\frac{2x^2\sigma^2_0}{\tilde\sigma^4}N}
  \left[1+2\,C_0^2\cos\left(N\theta+\frac{2 N x_0}{\tilde\sigma^2}x\right)\right].
\end{equation*}
Notice that the above probability
depends on $\theta$ only for states with non-negligible NOON component, i.e. non-zero $C_0$ (and thus $C_N$) \cite{bach}. 

\begin{figure}
  \includegraphics[scale=.36,clip]{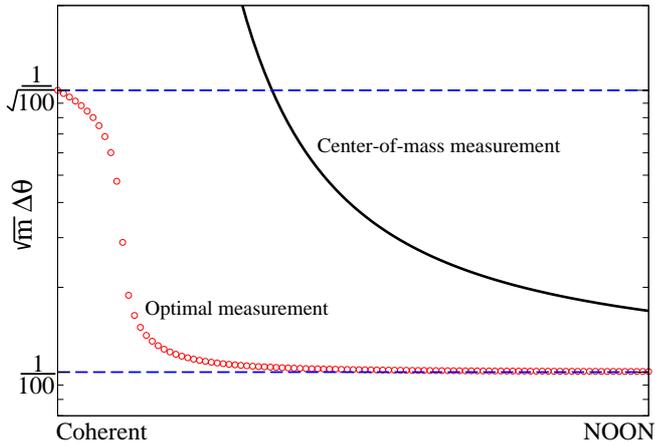}
  \caption
      {
        (color online). The sensitivity $\sqrt m\Delta\theta$ (black solid line)  for $N=100$
        calculated with Eq.(\ref{fish_cm}) as a function of $|\psi_{in}\rangle\in\mathcal{A}$. The sub shot-noise
	sensitivity is reached for all states with non-negligible NOON component (see Eq.(\ref{fish_cm}) for details).
        The values of $\sqrt m\Delta\theta_{SN}$ and $\sqrt m\Delta\theta_{HL}$ are
        denoted by the upper and lower dashed blue lines, respectively. The optimal sensitivity, given by the inverse of
        the QFI, is drawn with the red open circles. 
      }
      \label{cm}
\end{figure}

When $p_{cm}(x|\theta)$ is known,
the phase can be estimated using the MLE analysis described in the previous section. 
The sensitivity is given by the inverse of the FI which can be calculated analytically using the definition
$F_{cm}=m
\int\limits_{-\infty}^\infty dx\frac{1}{p_{cm}(x|\theta)}\left(\frac{\partial}{\partial\theta}p_{cm}(x|\theta)\right)^2$.
We obtain
\begin{equation}\label{fish_cm}
  F_{cm}=mN^2\left[1-\sqrt{1-2\,C_0^2}\right].
\end{equation}
In Fig.\ref{cm} we plot the sensitivity calculated by the inverse of the FI (\ref{fish_cm}) as
a function of $|\psi_{in}\rangle\in\mathcal{A}$. 
The estimation through the center of mass is not optimal, but the sensitivity can still be better than shot-noise,
and tends to $\Delta\theta_{HL}$ for $|\psi_{in}\rangle\rightarrow$ NOON.
We underline that, although the calibration stage is not as difficult as in the case of high-order correlations, 
phase estimation based on the center of mass measurement requires
detection of {\it all} $N$ atoms. It can be demonstrated that 
if just one particle is missed,  the sub shot-noise sensitivity is inevitably lost. 

{\it Conclusions}.
We analyzed different phase estimation strategies based on the position
measurement of atoms released from a double-well trap. 
We demonstrated that 
the fit to the density gives a sensitivity limited by the shot-noise. 
This bound can be overcome by phase estimation with correlation functions of the order of at least $\sqrt N$, 
while the $N$-th order correlation function provides an {\it optimal} estimation strategy. 
We also showed that the measurement of the position of the center of mass 
gives sub shot-noise sensitivity for all states with non-negligible NOON component. 
Both the protocol involving high-order correlations and the one based on the center of mass position
are difficult to implement. 
The difficulty in achieving sub shot-noise sensitivity can be attributed to the fact that, 
after formation of the interference 
pattern, the two spatial modes, corresponding to the arms of the interferometer, cannot be distinguished
and the useful information about the correlations between these modes is unavailable. To beat the shot-noise limit,
one has to determine the correlations between the particles. 
These correlations are difficult to extract from the experimental data,
thus sub shot-noise sensitivity
with two interfering BECs might prove challenging.

{\it Acknowledgments}. We acknowledge fruitful discussions with Julian Grond, whose constructive 
remarks lead to major improvement of the manuscript.


\begin{thebibliography}{40}
\bibitem{nota_corr}
  {
    The $N$-th order correlation function reads $G_N(\vec x_N,\theta)=\langle\psi_{out}|
    \hat\Psi^\dagger(x_1,t)\ldots\hat\Psi^\dagger(x_N,t)\hat\Psi(x_N,t)\ldots\hat\Psi(x_1,t)|\psi_{out}\rangle.$
    Here, $|\psi_{out}\rangle=e^{-i\theta\hat J_z}|\psi_{in}\rangle$ 
    is the state of the system after phase acquisition.
  }
\bibitem{nota_cn}
  {
    This assumption is satisfied for any ground state of a double-well system with either attractive or positive
    interactions.
  }
\bibitem{nota_MLE}{Other estimation strategies, like the Bayesian analysis, could be used.}
\bibitem{nota_qfi}
  {
    The value of the QFI is given by  $4 m$ times the variance (in the input state)
    of the operator generating the interferometric transformation.
    In the case of the phase-shift, the generator is $\hat J_z$ thus $F_Q=m\cdot 4\Delta^2\hat J_z$. 
  }
\end{thebibliography}
\end{document}